\documentclass[amsfonts,amsmath,prd,twocolumn,nofootinbib,a4paper,eqsecnum,superscriptaddress]{revtex4-2}
\newcommand{\beq}{ \begin{equation}}
\newcommand{\eeq}{\end{equation}}

\newcommand{\bsp}{\begin{split}}

\usepackage{epsfig,bbm,cancel}
\usepackage[normalem]{ulem}
\usepackage[colorlinks=true]{hyperref}
\usepackage{latexsym}
\usepackage[utf8]{inputenc}
\usepackage{amsmath}
\usepackage{amsfonts}
\usepackage{amssymb}
\usepackage{booktabs}
\usepackage{array}
\usepackage{tabularx}
\usepackage{multirow}
\usepackage{longtable}
\usepackage{ragged2e}
\usepackage{braket,booktabs}
\usepackage{braket,amsmath}
\usepackage{xcolor}
\usepackage{graphicx}
\usepackage{orcidlink} 

\begin{document}

\title{Axial Perturbations of Extended Tolman III in the New $f(R,T)$ Theory}

\author{B. N. Jayawiguna \orcidlink{0000-0001-5920-8701}}
\email{nugrahabyon312@gmail.com}
\author{G. A. Sabian \orcidlink{0009-0002-4175-3626}}
\email{gavin.athaya@ui.ac.id}
\author{H. S. Ramadhan \orcidlink{0000-0003-0727-738X}}
\email{hramad@sci.ui.ac.id}
\author{A. Sulaksono \orcidlink{0000-0002-1493-5013}}
\email{anto.sulaksono@sci.ui.ac.id}
\affiliation{Departemen Fisika, FMIPA, Universitas Indonesia, Depok, 16424, Indonesia.}

\begin{abstract}
We investigate compact objects using the extended Tolman III equation of state in the framework of $f(R,T)=R+2\chi T$ gravity. In the matter sector of the Lagrangian density, we employ a new formalism based on the evaluation of the second-order derivative of the matter Lagrangian with respect to the metric \cite{Jayawiguna:2026das}. We analyze the corresponding observational signatures through axial gravitational perturbations including stellar stability and gravitational echo. We find that the effective potential in $f(R,T)$ gravity is wider and deeper than its counterpart in general relativity (GR). The quasinormal-mode (QNM) frequencies are calculated using the Bohr--Sommerfeld (BS) fitting and both the lower- and higher-order Wentzel--Kramers--Brillouin (WKB) approximations. For the ultracompact configurations considered, trapped modes can tunnel through the potential barrier, giving rise to gravitational-wave echoes. The imaginary parts of the QNM frequencies are found to be negative, $\Im(\omega_n M)<0$, indicating that the perturbations decay exponentially with time. Our results show that the echo train in GR contains more echoes than that in $f(R,T)$ gravity. Finally, assuming that the stellar object is a neutron star and adopting a stellar radius consistent with measurements by the Neutron Star Interior Composition Explorer (NICER), we find that the $f(R,T)$ gravity model yields a characteristic frequency consistent with the estimated frequency associated with GW170817 for $\beta= 4.663\times10^{-1}$ and $\chi=0.96\pi$, whereas the corresponding GR model does not exhibit comparable agreement. 

\vspace{0.2 cm}

\end{abstract}

\maketitle
\thispagestyle{empty}
\setcounter{page}{1}

\section{Introduction}

Einstein's general theory of relativity (GR) is the fundamental theory of gravity in the relativistic regime. It provides a powerful framework for describing gravitational phenomena and has been extensively tested across a broad range of physical scales and energy regimes (see, e.g., \cite{Shankar2022,Berti:2015itd,Blazquez-Salcedo:2022dxh}). Nevertheless, the application of GR to extreme physical regimes remains an active area of research, particularly in the ultraviolet (UV) regime, where strong-field gravity and
high-energy physics become relevant, and in the infrared (IR) regime, which is associated with large-scale cosmological phenomena. In the UV regime, modifications to GR or additional physical contributions may be required to investigate the internal structure and stability of compact objects, such as neutron stars (NSs), quark stars (QSs), and white dwarfs (WDs), as well as the singularity problem in black holes (BHs) \cite{delaCruz-Dombriz:2009pzc,Cembranos:2011sr,
Sheykhi:2012zz,Tang:2019qiy,Khodadi:2020cht,Khodadi:2022xtl}. Modified gravity theories introduce additional parameters or effective geometrical structures that can alter the properties of compact stars and black holes, potentially providing  insights into gravitational phenomena beyond the standard predictions of GR.

NSs with realistic equations of state (EoSs) have masses ranging from approximately $1.4$ to $2M_{\odot}$ and radii typically within the range of 10–20 km~\cite{Clark:2002db,Antoniadis:2013pzd,Miller:2021qha}. Therefore, NSs are classified as compact objects. Moreover, the nature of the massive secondary object detected by LIGO and Virgo in the GW190814 event, with a mass of $2.50–2.67 M_{\odot}$, remains uncertain, as its physical properties do not clearly establish whether it is a neutron star or a black hole~\cite{LIGOScientific:2020zkf}. Studies~\cite{Yazadjiev:2014cza,Capozziello:2015yza,AparicioResco:2016xcm,Astashenok:2017dpo,Astashenok:2018iav,Feola:2019zqg,Jayawiguna:2022ftj,Jayawiguna:2023vvw,Jayawiguna:2025edl} have generally shown that the mass–radius profiles of NSs can shift toward higher masses as the strength of gravitational modifications increases. Thus, modified gravity may provide an alternative framework for accommodating compact objects with higher maximum masses.

For WDs, the Chandrasekhar limit, approximately $1.4 M_{\odot}$, is commonly regarded as the maximum mass in general relativity (GR)~\cite{shapiro,lauffer,woosley,Chandrasekhar:1931ftj,Chandrasekhar:1931ih}. However, recent studies have suggested the existence of super-Chandrasekhar WDs with masses of $2.1–2.8\,M_{\odot}$~\cite{Hicken:2007ap,Hillebrandt:2000ga,Khokhlov1993}, potentially associated with overluminous Type Ia supernovae~\cite{SNLS:2006ics,Filippenko:1992wda,Taubenberger:2007dt,Turatto:1998eq,Silverman2013,Yamanaka2009,Scalzo2010}. These observations motivate the study in Ref.~\cite{Priyobarta:2026dip}, which constructs WD models within several classes of $f(R)$ gravity theories. Their results show that modified gravity can yield maximum WD masses exceeding the GR limit, a possibility referred to as the super-Chandrasekhar limit.

On the other hand, QSs are intriguing because they represent a hypothetical third class of compact stellar remnants, potentially denser than neutron stars but not necessarily forming black holes. They provide a unique laboratory for testing fundamental physics at extreme densities. QSs are hypothesized to consist of strange quark matter, a deconfined state of quarks that may represent the most stable form of matter in the universe. For recent reviews of quark stars and their corresponding equations of state (EoSs), see Refs.~\cite{Kumari2021,Danarianto2025,Zhang2024}. Pretel \textit{et al.}~\cite{Pretel:2022plg} studied anisotropic quark stars in \(f(R)=R^{1+\epsilon}\) gravity, with \(\epsilon \ll 1\), using the MIT bag model EoS. The nature of certain compact objects remains an open question, with possible interpretations including low-mass black holes, rapidly rotating neutron stars, quark stars, and hybrid stars; no consensus has been reached~\cite{Annala2020}. See also Ref.~\cite{Prasetyo2021} for further details.

In the (IR) cosmological regime, modifications to general relativity (GR) have been proposed to explain the late-time accelerated expansion of the universe~\cite{WMAP:2006bqn,WMAP:2008lyn,Komatsu_2011,Li:2011sd,SupernovaCosmologyProject:1998vns,SupernovaSearchTeam:1998fmf,SupernovaCosmologyProject:2003dcn}. This approach gained significant attention following the pioneering work of Starobinsky~\cite{Starobinsky:1980te}. In Ref.~\cite{Carroll:2003wy}, a model of the form $f(R)=R-\mu^4/R$ was proposed to explain the accelerated expansion of the universe, where $\mu$ is a mass scale approximately equal to the present-day Hubble parameter, $\mu \approx 10^{-33}~\text{eV}$. The authors showed that the dark-energy-like contribution can emerge from the effective gravitational dynamics of the geometry, without requiring an additional cosmological constant. Another important feature of $f(R)$ gravity is its equivalence to a scalar–tensor theory. Specifically, metric $f(R)$ gravity can be reformulated as a Brans–Dicke-type theory~\cite{PhysRev.124.925}, in which the gravitational sector includes a scalar degree of freedom nonminimally coupled to gravity. Related studies have also explored such scalar–tensor frameworks in the context of black holes (BHs)~\cite{Rodrigues:2015ayd,Rodrigues:2016fym,Hollenstein:2008hp}.

Moreover, the $f(R)$ gravity framework can be further generalized by introducing an explicit coupling between curvature and the matter Lagrangian in the gravitational action. Studies~\cite{Bertolami:2007gv,Bertolami:2008ab} have shown that such curvature–matter couplings can give rise to an additional force and lead to the non-conservation of the matter energy-momentum tensor, potentially providing a mechanism for explaining cosmic acceleration~\cite{Bisabr:2012tg}. Subsequently, Harko \textit{et al.}~\cite{Harko:2011kv} proposed a generalized class of theories, known as $f(R,T)$ gravity, in which the action depends on both the Ricci scalar \(R\) and the trace $T$ of the energy-momentum tensor. A key feature of $f(R,T)$ gravity is that the field equations can depend on the second functional derivative of the matter Lagrangian with respect to the metric tensor, $\delta^2 \mathcal{L}_{m}/\delta g^{\mu\nu} \delta g^{\alpha\beta}$, which must be specified in the variational procedure. In GR, different choices of the matter Lagrangian, such as $\mathcal{L}_{m}=p$ and $\mathcal{L}_{m}=-\rho$, can yield equivalent Einstein field equations for a perfect fluid under appropriate assumptions. However, in theories involving explicit curvature–matter couplings, these choices can lead to different field equations. This ambiguity arises in the variational procedure because the standard fluid variables, including the four-velocity $u_{\mu}$ and the metric tensor $g_{\mu\nu}$, are treated as independent degrees of freedom~\cite{Akarsu:2023lre}. In Ref.~\cite{Jayawiguna:2026das}, the authors address this issue by explicitly incorporating the normalization condition of the four-velocity into the variational procedure. They demonstrate that, within this approach, the resulting field equations become independent of the specific choice of the matter Lagrangian $\mathcal{L}_{m}$. In the present work, we adopt the ``new formalism'' developed in Ref.~\cite{Jayawiguna:2026das} and apply it to the extended Tolman III solution, which describes a constant-density star (CDS), within the framework of $f(R,T)$ gravity.

An important phenomenological aspect of astrophysics is the study of non-radial oscillations and gravitational-wave (GW) events. The detections of GW150914, GW170817, and subsequent compact-object mergers have firmly established the existence of GWs, opening new avenues for both observational and theoretical investigations~\cite{LIGOScientific:2016aoc,LIGOScientific:2017vwq,LIGOScientific:2021qlt}. The compactness, $\beta\equiv M/R$, plays a central role in characterizing compact objects~\cite{CP2019}. The black-hole limit corresponds to $\beta=0.5$, whereas ultracompact objects (UCOs), with compactness in the range $0.33\leq \beta \leq \beta_{\textrm{B}}$, can possess a photon sphere and potentially produce GW echoes due to the presence of an effective potential well. Here, $\beta_{\textrm{B}}$ denotes the Buchdahl-limit compactness. In GR, gravitational perturbations can be decomposed into polar (even-parity) and axial (odd-parity) modes. The axial sector, which describes pure spacetime perturbations, is governed by a Regge--Wheeler form with an effective potential. Solving this equation yields complex quasinormal-mode (QNM) frequencies, whose imaginary parts describe the damping of the oscillations that decays in time. Previous studies have shown that, in ultracompact configurations, the effective potential can trap GWs and give rise to resonant modes~\cite{Urbano:2018nrs}. In \cite{Yu:2026vbd}, recent study shows that the spacetime mode was found to be stable in anisotropic cases.

In this work, as discussed in the previous section, we adopt the formalism developed in Ref.~\cite{Jayawiguna:2026das} for $f(R,T)$ gravity to investigate the speed-of-sound profile, energy conditions, and axial perturbation properties of constant-density stars. It is important to note that a closely related study, excluding axial perturbations, was conducted by Pappas \textit{et al.}~\cite{Pappas:2022gtt} using different prescriptions. We compare our results with their findings and the corresponding GR results across the parameter space. Furthermore, we extend the analysis to the perturbation sector by deriving the effective potential, calculating quasinormal-mode (QNM) frequencies, investigating gravitational-wave echoes, and estimating characteristic frequencies that may be associated with GW170817. Although the Tolman III solution (also known as the Schwarzschild interior solution or Schwarzschild star) assumes a uniform energy-density profile and is therefore less physically realistic than models such as Tolman VII, whose quadratically decreasing density profile better approximates realistic NS EoSs, it offers several advantages for the present analysis. First, the metric functions, pressure, and mass function can be written analytically. Second, the stellar structure is specified by the compactness, which can therefore be varied across the entire range, up to the Buchdahl limit, $\beta_{\textrm{B}}$, at which the central pressure diverges. In particular, this range includes the ultracompact regime \cite{Cardoso:2016oxy}, where the effective potential develops a local minimum and supports trapped QNMs with the negative imaginary part. Nevertheless, the constant density star (CDS) model has an infinite speed of sound, a pathology that, as we will see later, is resolved in $f(R,T)$ gravity. For a more realistic, yet analytically tractable stellar model, the Tolman VII solution and its extensive literatures are discussed in Refs.~\cite{Tolman:1939jz,Lattimer:2000nx,Jiang:2019vmf}.

This paper is structured as follows: Section \eqref{section2} discusses the constant density star model under the influence of $f(R,T)$ gravity, equations of motion, and the analytic expressions of the metric tensor. In Section \eqref{section3}, we present the parametric space to constrain the realistic region by applying the subluminal speed of sound and  modified energy conditions. This region could be important to analyze the axial gravitational perturbation which we will present in Section \eqref{section4}. Lastly, we conclude our results in Section \eqref{section5}.

\section{Formalism}
\label{section2}

The general action involving matter in the gravity part \cite{Harko:2011kv,Pappas:2022gtt} can be written as 
\begin{equation}
S = \int d^{4}x \sqrt{-g} \bigg[ \frac{f(R,T)}{16\pi} + \mathcal{L}_{m} \bigg],
\end{equation}
where $g$ is the determinant of the metric $g_{\mu\nu}$, and the quantity $\mathcal{L}_{m}$ is the Lagrangian in the matter part. By varying the latter action with respect to the metric, we can obtain the field equations in $f(R,T)$ gravity
\begin{equation}
f_{R}R_{\mu\nu}-\frac{1}{2}f g_{\mu\nu} + (g_{\mu\nu}\Box-\nabla_{\mu}\nabla_{\nu})f_{R} = 8\pi T_{\mu\nu} - f_{T}(T_{\mu\nu}+\Theta_{\mu\nu}),
\end{equation}
where we denote $f_{R} $ and $f_{T}$ as the partial derivatives of $f(R,T)$ with respect to $R$ and $T$, respectively. The quantity $\Theta_{\mu\nu}$ is defined as
\begin{eqnarray}
\label{Theta}
\Theta_{\mu\nu} &=& g^{\alpha\beta} \delta T_{\alpha\beta}/\delta g^{\mu\nu},\nonumber \\ &=&-2T_{\mu\nu}+g_{\mu\nu}\mathcal{L}_{m}-2g^{\alpha\beta} \frac{\partial^2\mathcal{L}_{m}}{\partial g^{\mu\nu}\partial g^{\alpha\beta}},
\end{eqnarray}
where the second derivative in the curvature-matter term exists in the last term. The energy-momentum tensor $T_{\mu\nu}$ is defined as 
\begin{equation}
T_{\mu\nu} = \frac{-2~\delta(\sqrt{-g} \mathcal{L}_{m})}{\sqrt{-g}~\delta g^{\mu\nu}}.
\end{equation}
However, we consider only interior star models with an isotropic perfect fluid
\begin{equation}
\label{isopf}
T_{\mu\nu} = (\rho+p)u_{\mu}u_{\nu}+pg_{\mu\nu},
\end{equation}
where $\rho(r)$ and $p(r)$ are the density and pressure of the star as a function of the radial coordinate, while $u^{\mu}$ is the 4-velocity of the fluid with the normalization $u^{\mu}u_{\mu}=-1.$

In order to further solve the equations of motion, we adopt the special linear class of $f(R,T)$ gravity considered in \cite{Pappas:2022gtt}, in which no mixing terms between the Ricci scalar and the trace of the energy-momentum tensor are present:
\begin{equation}
f(R,T) = R + 2\chi T.
\end{equation}
Consequently, the field equations simplify to
\begin{equation}
G_{\mu\nu}
= 8\pi T_{\mu\nu}
+ \chi g_{\mu\nu} T
- 2\chi (T_{\mu\nu}+\Theta_{\mu\nu}).
\end{equation}
Following the formalism developed in \cite{Jayawiguna:2026das}, we adopt an alternative prescription for evaluating $\mathcal{L}_m$ and its second variation with respect to the metric.%
\footnote{In \cite{Pappas:2022gtt}, an extended Tolman III solution is studied in $f(R,T)$ gravity. Following \cite{Harko:2011kv}, the authors adopt $\mathcal{L}_m=p$, for which the second derivative of $\mathcal{L}_m$ vanishes. Here, we employ a different prescription, which leads to different results.}
The second variations appearing in this formalism can be expressed compactly as \cite{Jayawiguna:2026das}
\begin{eqnarray}
\frac{\delta^2 p}
{\delta g^{\mu\nu}\delta g^{\alpha\beta}}
&=&
-\frac{\rho+p}{4}u_\mu u_\nu
\left(g_{\alpha\beta}+2u_\alpha u_\beta\right),
\\
\frac{\delta^2 \rho}
{\delta g^{\mu\nu}\delta g^{\alpha\beta}}
&=&
-\frac{\delta^2 p}
{\delta g^{\mu\nu}\delta g^{\alpha\beta}}
+\frac{\rho+p}{4}g_{\alpha\beta}g_{\mu\nu}
-\frac{\rho+p}{2}g_{\mu\alpha}g_{\nu\beta}.\nonumber \\ 
\end{eqnarray}
With these expressions, one obtains
\begin{equation}
\Theta_{\mu\nu}(\mathcal{L}_m=p)
=
\Theta_{\mu\nu}(\mathcal{L}_m=-\rho).
\end{equation}
Substituting this result into Eq.~\eqref{Theta} yields
$\Theta_{\mu\nu}=-T_{\mu\nu}$, and the field equations reduce to
\begin{equation}
\label{efe}
G_{\mu\nu}
= 8\pi T_{\mu\nu}+\chi g_{\mu\nu}T.
\end{equation}
Thus, the effect of the linear $T$-dependent term remains explicitly encoded in the last term on the right-hand side of Eq.~\eqref{efe}. 

With the static and spherically symmetric ansatz metric, the line element can be written as 
\begin{equation}
\label{ansatz}
ds^2 = -e^{\nu} dt^2 + e^{\lambda} dr^2 + r^2 d\Omega^2,
\end{equation}
where the variables in the time and radial components are functions of the radial coordinate only and $d\Omega^2 = d\theta^2 + \sin^2\theta ~d\phi^2.$ The Einstein field equation now reads
\begin{eqnarray}
\label{00}
\frac{1}{r^2} \frac{d}{dr}(re^{-\lambda}) - \frac{1}{r^2} &=& 8\pi\rho +\chi(\rho-3p), \nonumber \\ \\ e^{-\lambda} \left( \frac{\nu'}{r}+\frac{1}{r^2} \right) - \frac{1}{r^2} &=& 8\pi p +\chi(3p-\rho), \nonumber \\ \\ \frac{e^{-\lambda}}{4r} [2\nu'+r\nu'^2-\lambda'(2+r\nu')+2r\nu''] &=& 8\pi p + \chi (3p-\rho).\nonumber \\
\end{eqnarray}
Observe that the contribution of the linear class model of $f(R,T)$ is located at the second term in the right-hand side of the equation. When we set $\chi$ to 0, the equation comes back to the well-known Einstein equation in standard GR. However, the last two terms above can be reduced to the following equation for obtaining the $\nu(r)$ profile 
\begin{equation}
r^2 (2\nu''+\nu'^2-\nu'\lambda')-2r(\nu'+\lambda')+4(e^{\lambda}-1)=0.
\end{equation}
Next, we define the radial component metric to be 
\begin{equation}
e^{-\lambda} = 1-\frac{2m(r)}{r}.
\end{equation}
Equation \eqref{00} now reads
\begin{equation}
\label{mass}
\frac{dm}{dr} = 4\pi r^2 \rho + \frac{\chi}{2} (\rho-3p) r^2,
\end{equation}
and the TOV equation can be evaluated as
\begin{eqnarray}
\label{tov}
\frac{dp}{dr} &=& -\frac{8\pi(\rho+p)}{8\pi+3\chi} \bigg[\frac{m+4\pi r^3 p + \frac{\chi}{2} r^3 (3p-\rho) }{r(r-2m)} \bigg]  \nonumber \\ &&+  \frac{\chi}{8\pi+3 \chi} \frac{d\rho}{dr}.
\end{eqnarray}
The GR case can be obtained by setting $\chi=0.$ Therefore, the system equations \eqref{mass} and \eqref{tov} can be solved by imposing the boundary conditions, $m(0)=0$, and $p(0)=p_{c}$, with $p_{c}$ defined as the central pressure of the star. A fixed value of $p_{c}$ yields the corresponding mass $M$ and radius $R$ profiles. Thus, by integrating simultaneously with some intervals of $p_{c}$, we can get the mass-radius profile of the star for a chosen equation of state (EoS).

In this case, we consider the Tolman III model (also known as the Schwarzschild star) extended in $f(R,T)$ gravity. Following the approach from \cite{Tolman:1939jz}, we obtain the exact solution for the constant density star (CDS) in the context of $f(R,T)$ gravity
\begin{widetext}
\begin{eqnarray}
g_{tt}(x) &=& \frac{16\pi^2}{(8\pi+\chi)^2} \left[3\sqrt{1-2\beta} - \sqrt{1-2\beta x^2} \left( 1- \frac{\chi}{4\pi} \right) \right]^2,\\
g_{rr} &=& (1-2\beta x^2)^{-1},\\ \rho(x) &=& \frac{3\beta}{4\pi+2\chi} \left[ \frac{3\sqrt{1-2\beta}-\sqrt{1-2\beta x^2}+\frac{\chi}{4\pi}(\sqrt{1-2\beta x^2}+3\sqrt{1-2\beta}) }{3\sqrt{1-2\beta}-\sqrt{1-2\beta x^2}+\frac{\chi}{4\pi} \sqrt{1-2\beta x^2}} \right], \\ \label{pressure} p(x)&=& \frac{3\beta\left(1-\frac{\chi}{4\pi} \right)}{(4\pi+2\chi)} \left[\frac{\sqrt{1-2\beta x^2}-\sqrt{1-2\beta}}{3\sqrt{1-2\beta}-\sqrt{1-2\beta x^2}+\frac{\chi}{4\pi} \sqrt{1-2\beta x^2}}\right],
\end{eqnarray}
\end{widetext}
where $\beta \equiv M/R$ is the compactness and we have used the dimensionless radial coordinate $x\equiv r/R$, with $x\in[0,1]$. The results above constitute a complete solution of the constant density star under the influence of $f(R,T)$ gravity. The GR case can be found by setting the free parameter $\chi$ to zero. Observe that the density of the star, in this modified gravity, is no longer constant. Another aspect that we want to analyze is how large the gap is between linear $f(R,T)$ gravity and standard GR. This can be obtained by considering the speed of sound and the energy condition, which will be discussed in the next section.

\section{parametric space in $f(R,T)$ gravity}
\label{section3}

The complete solutions depend on the compactness and the dimensionless parameter. Although both can take arbitrary values, we restrict them to a specific range to analyze the relevant quantities in the axial gravitational perturbation sector.

\subsection{Speed of sound and maximum compactness based on finiteness pressure}

First, the speed of sound expression is given by 
\begin{equation}
c_s^2=\frac{dp}{d\rho} =\frac{1}{3} + \frac{8\pi}{3\chi}.
\end{equation}
Unlike in the GR case, we can see that the profile above is no longer infinite. It is physically expected due to the radial dependence in the density of the star coming from the linear $f(R,T)$ model, and the latter expression is independent of the compactness $\beta$. The (subluminal) speed of sound can be valid in the region, $0<c_s^2<1,$ and therefore we can restrict the dimensionless parameter to be $\chi\in[-8\pi,4\pi]$, where the starting (final) point interval comes from the $c_s^2>0$ $(c_s^2<1)$ condition. 

The Buchdahl limit in linear $f(R,T)$ gravity can be seen from the pressure \eqref{pressure} near the center of the star. The highest compactness is obtained when the central pressure is very high or the denominator is zero. Therefore, we get
\begin{equation}
\label{betabuchdal}
\beta_{\textrm{B}} = \frac{128 \pi^2 + 8\pi \chi -\chi^2}{288\pi^2}.
\end{equation}
Note that in the small regime, $\chi\ll1,$ the equation above can be expanded as 
\begin{equation}
\beta_{\textrm{B}} \approx \frac{4}{9} + \frac{\chi}{36\pi} - \frac{\chi^2}{288\pi^2} + \mathcal{O}(\chi^3),
\end{equation}
and the GR Buchdahl limit is recovered in the zeroth order of the expansion. Next, we study the energy condition to ensure the realistic condition for the star. Note that the energy conditions in the $f(R)$ family are different from GR \cite{Santos:2007bs,Arora:2024xrp,Wang:2012rw,Atazadeh:2013cz}, since they contain the d'Alembertian in the field equation. In the standard GR, when $f(R)=R,$ the d'Alembertian term is zero.

\subsection{Energy conditions}
Involving the matter Lagrangian in the gravity sector causes a deviation in the field equations with respect to GR. Therefore, the energy conditions in $f(R,T)$ gravity can be adopted within the original GR framework. We briefly review the method we used to obtain the energy conditions: the strong energy condition (SEC), the weak energy condition (WEC), the dominant energy condition (DEC), and the null energy condition (NEC). The origin theory of these latter profiles comes from the Raychaudhuri equation with the requirement that gravity be attractive. 

To begin with, let $u^{\mu}$ be the tangent vector field to a congruence of timelike geodesics in a spacetime manifold embedded with the metric $g_{\mu\nu}$. The equation reads \cite{Santos:2007bs}
\begin{equation}
\frac{d\theta}{d\tau} = -\frac{1}{3} \theta^2 - \sigma_{\mu\nu}\sigma^{\mu\nu} + \omega_{\mu\nu}\omega^{\mu\nu} - R_{\mu\nu}u^{\mu}u^{\nu},
\end{equation}
where $R_{\mu\nu}$ is the Ricci tensor, while the quantities $\theta$, $\sigma^{\mu\nu}$, and $\omega_{\mu\nu}$ are the expansion, shear, and rotation, respectively, related to the congruence defined by the vector field. In this context, we follow the condition for attractive gravity, therefore $d\theta/d\tau<0$ and $\omega_{\mu\nu}=0.$ Since the shear one is a spatial tensor, one has $\sigma^2 = \sigma_{\mu\nu}\sigma^{\mu\nu}\geq0$. The equation reduces to
\begin{eqnarray}
\label{sec}
R_{\mu\nu}u^{\mu}u^{\nu} &=& 8\pi \left(T_{\mu\nu} -\frac{1}{2} g_{\mu\nu} T \right)u^{\mu}u^{\nu}\geq0,\\ \label{nec} R_{\mu\nu}k^{\mu}k^{\nu} &=& 8\pi T_{\mu\nu} k^{\mu} k^{\nu} \geq 0.
\end{eqnarray}
Equations \eqref{sec} and \eqref{nec} are known as the SEC and NEC conditions. For the isotropic perfect fluid \eqref{isopf}, both equations reduce to $\rho+3p\geq 0,$ and $\rho+p\geq 0,$ respectively. Another energy condition, namely the WEC, mandates the positivity of the energy density for any observer at any point, indicating $\rho \geq 0$ and $\rho+p\geq 0$. The requirement that energy cannot propagate faster than the speed of light gives rise to an additional condition, namely $\rho-p \geq 0$, along with the condition $\rho \geq 0$, which is called the DEC. 

In the linear $f(R,T)$ gravity, the field equations can be recast as effective equations
\begin{equation}
R_{\mu\nu} =  \left( \tilde{\mathcal{T}}_{\mu\nu} - \frac{1}{2}g_{\mu\nu}\tilde{\mathcal{T}}\right),
\end{equation}
where
\begin{eqnarray}
\tilde{\mathcal{T}} &=&\frac{1}{  f_{R}}(8\pi\tilde{T}+f-Rf_{R}-3\Box f_{R}),\\ \tilde{\mathcal{T}}_{\mu\nu} &=& \frac{1}{ f_{R}}[8\pi \tilde{T}_{\mu\nu}+(\nabla_{\mu}\nabla_{\nu}-g_{\mu\nu}\Box)f_{R}]. 
\end{eqnarray}
The SEC equation \eqref{sec}, after some algebra, can be written as
\begin{eqnarray}
&& 8\pi(\rho+3p) + f-R f_{R} - \nu'e^{-\lambda} R' f_{RR} \nonumber \\ &&- e^{-\lambda} \bigg[ \left( \frac{2}{r} + \frac{\nu'-\lambda'}{2} \right)R' f_{RR} + R'' f_{RR} + R'^2 f_{RRR}   \bigg] \geq0. \nonumber \\ 
\end{eqnarray}

The NEC equations \eqref{nec} after some algebra, can be written as 
\begin{equation}
8\pi (\rho+p) + e^{-\lambda} \bigg[ R'' f_{RR} + R'^2 f_{RRR} - \frac{(\nu'+\lambda')}{2}R' f_{RR}  \bigg] \geq0,
\end{equation}
and the DEC reads 
\begin{eqnarray}
&&  e^{-\lambda} \bigg\lbrace \left[ \frac{4}{r} - \frac{(\lambda'-\nu')}{2} \right] R' f_{RR} + R''f_{RR} +R'^2 f_{RRR}  \bigg\rbrace \nonumber \\ &&+ 8\pi(\rho-p) -f+R f_{R} + \geq0.
\end{eqnarray}
Since the metric tensor, density, and the pressure are analytics, the exact solutions of the SEC, NEC, and DEC, respectively, can be expressed as 
\begin{widetext}
\begin{eqnarray}
\textbf{SEC}:&~~~~&   \frac{12 \beta  (\chi -4 \pi ) \sqrt{1-2 \beta 
		x^2}}{4 \pi  \left(\sqrt{1-2 \beta  x^2}-3
	\sqrt{1-2 \beta }\right)-\chi  \sqrt{1-2 \beta 
		x^2}}\geq 0,\\ \textbf{NEC}:&~~~~&   \frac{48 \pi  \sqrt{1-2 \beta } \beta }{12 \pi 
	\sqrt{1-2 \beta }+\chi  \sqrt{1-2 \beta  x^2}-4
	\pi  \sqrt{1-2 \beta  x^2}} \geq 0, \\ \textbf{DEC}:&~~~~&  \frac{12 \beta  \left(8 \pi  \sqrt{1-2 \beta }+\chi 
	\sqrt{1-2 \beta  x^2}-4 \pi  \sqrt{1-2 \beta 
		x^2}\right)}{12 \pi  \sqrt{1-2 \beta }+\chi 
	\sqrt{1-2 \beta  x^2}-4 \pi  \sqrt{1-2 \beta 
		x^2}} \geq 0.
\end{eqnarray}
\end{widetext}
Therefore, the maximum compactness within the DEC limit can be written as 
\begin{equation}
\label{betadec}
\beta_{\textrm{DEC}} = \frac{48\pi^2+ 8\pi\chi-\chi^2}{128\pi^2},
\end{equation}
where in the small $\chi$, the GR is recovered in the zeroth order, as expected
\begin{equation}
\beta_{\textrm{DEC}} \approx \frac{3}{8} + \frac{\chi}{16\pi} - \frac{\chi^2}{128\pi^2}+ \mathcal{O}(\chi^3).
\end{equation}

\begin{figure*}[t!]
	\centering
	\includegraphics[width=0.6\textwidth]{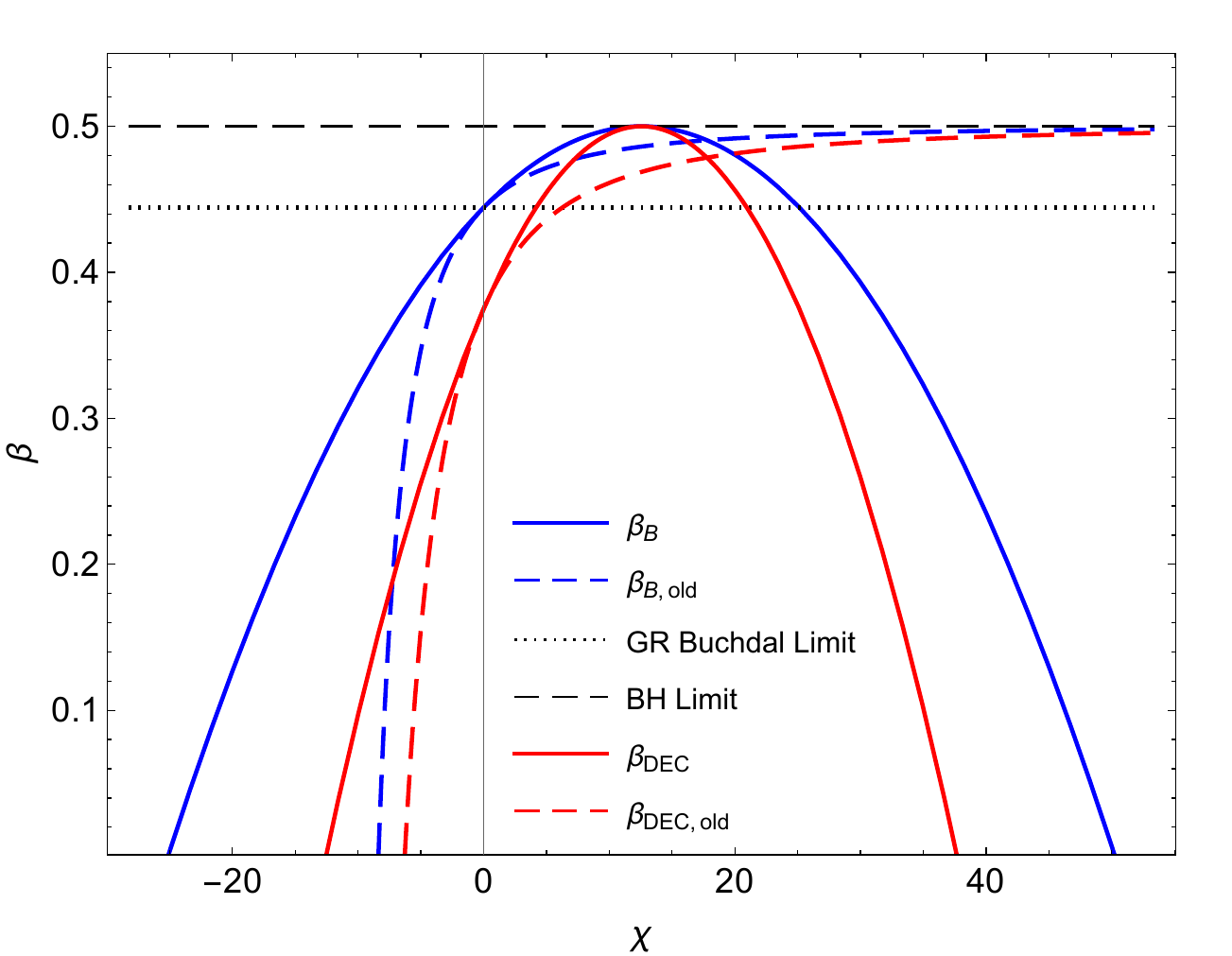}
	\caption{The parametric plot, $\chi-\beta$, in FRTG. Both blue and red solid lines are our work, while the dashed lines are the values from the work done by \cite{Pappas:2022gtt}. To compare, we also include the black dashed line and dotted line as the black hole (BH) limit and the GR Buchdahl limit, respectively.}
	\label{fig:betachi}
\end{figure*}

\begin{figure*}[t!]
	\centering
	\includegraphics[width=0.47\textwidth]{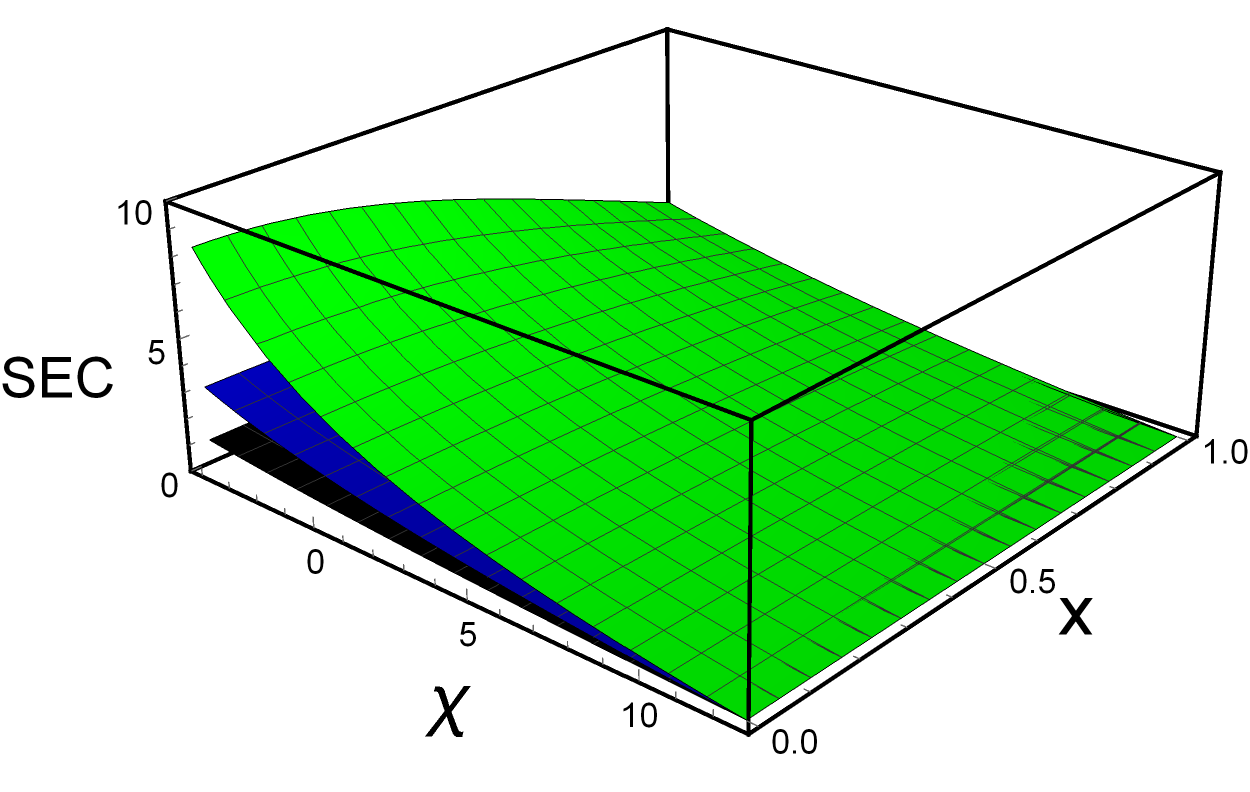}
	\includegraphics[width=0.47\textwidth]{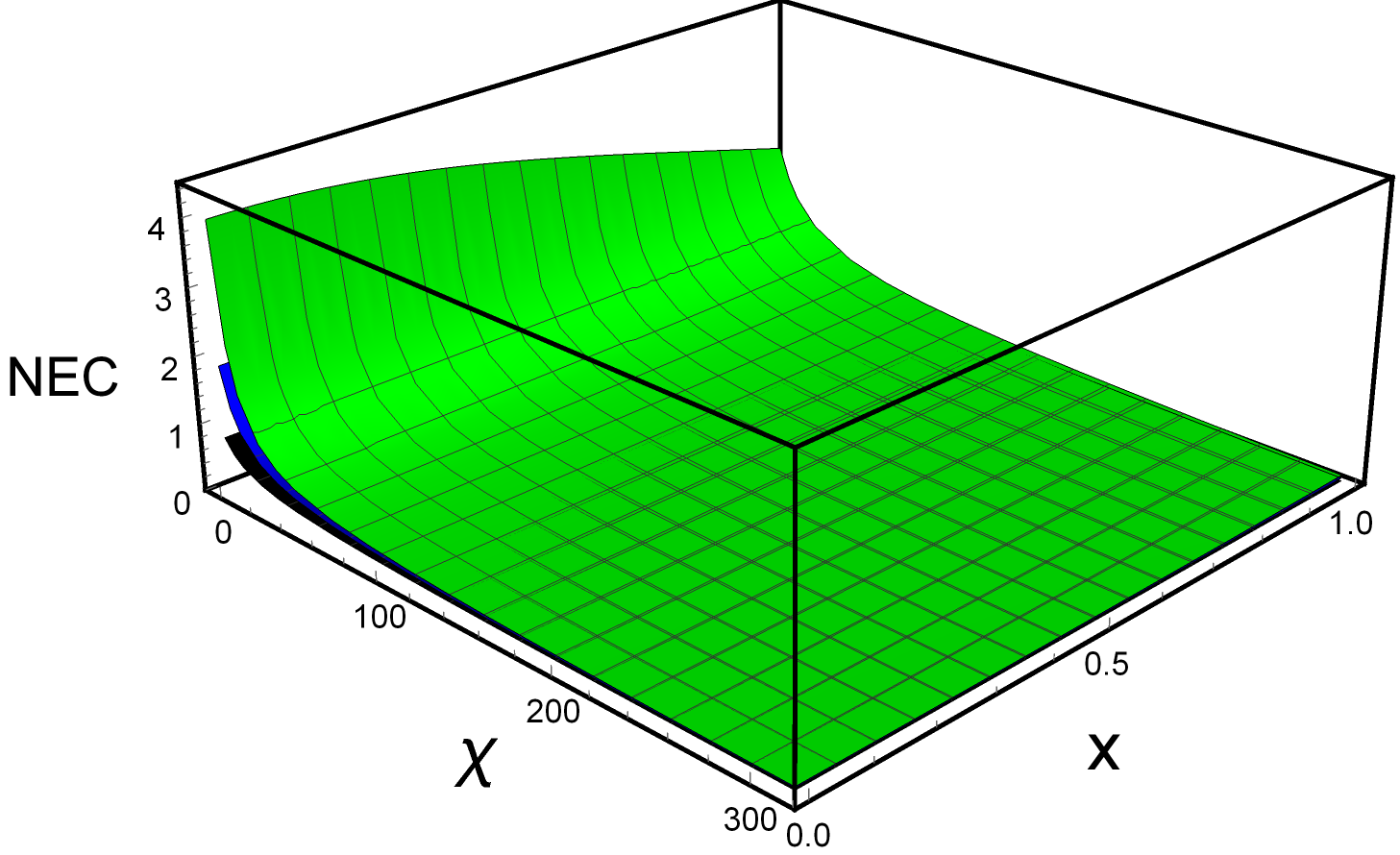}
	\caption{The energy conditions with three representative compactness values ($\beta = 0.1.~\beta=0.2,~\beta=0.3$).}
	\label{fig:ec}
\end{figure*}

\begin{table}[h!]
	\centering
	\begin{tabular}{||c|c|c|c|c||}
		\hline
		$\beta$ &  SEC & NEC  \\
		\hline
		0.1 & $\chi\in [-6.78\pi,4\pi]$ & $\chi\in [-6.7\pi,\infty]$ \\
		\hline
		0.2 & $\chi\in [-5.29\pi,4\pi]$ & $\chi\in [-5.3,\infty]$ \\
		\hline
		0.3 &$\chi\in [-3.61\pi,4\pi]$ &$\chi\in [-3.61\pi,\infty]$  \\
		\hline
		0.4 &$\chi\in [-1.3\pi,4\pi]$ &$\chi\in [-1.3\pi,\infty]$  \\
		\hline
		0.44 &$\chi\in [-0.15\pi,4\pi]$ & $\chi\in [-0.15\pi,\infty]$ \\
		\hline
	\end{tabular}
	\caption{The bounded range of the FRTG model's parameter $\chi$, from several representative compactness. The definition is $\chi \in[\chi_{\textrm{min}},\chi_{\textrm{max}}]$.}
	\label{table:1}
\end{table}

The expression for the Buchdahl limit and the DEC are presented in eqs.~\eqref{betabuchdal} and \eqref{betadec}. Fig.~\ref{fig:betachi} represents our numerical plot from the compactness condition over the $\chi$ value compared with the "old" one by the work from \cite{Pappas:2022gtt}. From the plot, we can see that our new formalism tends to have a bounded $\chi$ region, with $\chi_{\textrm{Buchdahl}}\in[-8\pi,16\pi]$, and $\chi_{\textrm{DEC}}\in[-4\pi,12\pi]$. The special properties occur in $\chi=4\pi,$ where the Buchdahl profile coincides with the DEC limit at the BH compactness limit. On the other hand, Fig.~\ref{fig:ec} displays a three-dimensional plot of the energy conditions of SEC and NEC, respectively, with respect to $x$ and $\chi$. Where $x$ is the radial coordinate that spans from the center ($x=0$) to the surface ($x=1$). Each of the profiles has a different range of $\chi$ based on the positivity condition we have to satisfy. The detailed value in the wider range of compactness are presented in the Table \ref{table:1}. It is shown that the region is bounded for the SEC and unbounded for the NEC. This can be proved by using the limit in each energy property. In the limit $\chi \gg 1,$ the SEC and NEC have the values 
\begin{eqnarray}
\textrm{SEC} \approx-12\beta,~~~~~\textrm{and}~~~~~\textrm{NEC} \approx 0.
\end{eqnarray}
The NEC has zero value in the high parameter, while the SEC needs to be handled with care, and we found that the upper bound is uniform.

In addition to the new realistic condition range of the star, the FRTG's parameter also extends the Buchdahl limit range compared to the GR counterpart. In the next section, we investigate the new maximum compactness value with a nonzero $\chi$ compared to the CDS Buchdahl limit in GR, $\beta=0.44$, and explore the observational imprints we can improve from these results.

\section{Gravitational (Axial) Perturbation; Effective potential, quasinormal mode, and gravitational wave echoes}
\label{section4}

In this section, we study the phenomenology appearing in our FRTG model. The effective potential and the quasinormal mode can be obtained from the perturbation method. In this case, we consider only the axial perturbations, which lead us to the wave equation with the Regge-Wheeler (RW) potential. For such perturbations, the very general metric can be written as\footnote{Important to note that this convention variable is only valid in this section.} \cite{Chandrasekhar:1985kt}
\begin{eqnarray}
ds^2 &=& e^{2\nu}dt^2 - e^{2\psi} (d\varphi-\omega dt-q_{2}dx^2-q_{3}dx^3)^2 \nonumber \\ && -e^{2\mu_{2}}(dx^{2})^2 - e^{2\mu_{3}}(dx^{3})^2,
\end{eqnarray}
where $\nu$, $\psi$, $\mu_{2}$, and $\mu_{3}$ denote the unperturbed configurations, while $\omega$, $q_{2}$, and $q_{3}$ denote the perturbation variables. The latter variables are functions of $t(=x^{0})$, $x^2(=r)$, and $x^3(=\theta)$. Our field equations have the form of Eq.~\eqref{efe}, and in the axial direction, the non-zero component of the Einstein field equation (in the tetrad frame) is [(1),(2)] and [(1),(3)]. The energy-momentum tensor with the latter components is zero. Therefore, the nonzero Ricci tensor can be written in a compact form
\begin{widetext}
\begin{eqnarray}
-R_{(1)(2)}&=& \frac{1}{2}e^{-2\psi-\nu-\mu_{3}} \bigg[  (e^{3\psi+\nu-\mu_{2}-\mu_{3}} Q_{32}),_{3}  - 
(e^{3\psi-\nu+\mu_{3}-\mu_{2}} Q_{02}),_{0}  \bigg]   , \\ \nonumber -R_{(1)(3)} &=&  \frac{1}{2}e^{-2\psi-\nu-\mu_{2}} \bigg[  (e^{3\psi+\nu-\mu_{3}-\mu_{2}} Q_{23}),_{2}  - 
(e^{3\psi-\nu+\mu_{2}-\mu_{3}} Q_{03}),_{0}  \bigg],
\end{eqnarray}
\end{widetext}
where 
\begin{equation}
Q_{AB}=q_{A,B}-q_{B,A}~~\textrm{and}~~ Q_{A0}=q_{A,0}-\omega_{,A}~~~(A,~B=2,~3).
\end{equation}
Hence, the perturbed Einstein field reads
\begin{eqnarray}
(e^{3\psi-\nu-\mu_{2}-\mu_{3}}Q_{23}),_{3} + e^{3\psi-\nu-\mu_{2}+\mu_{3}}Q_{02,0} &=& 0, \nonumber \\ (e^{3\psi-\nu-\mu_{2}-\mu_{3}}Q_{23}),_{2} - e^{3\psi-\nu+\mu_{2}-\mu_{3}}Q_{03,0} &=& 0,
\end{eqnarray}
With the aid of these two latter equations, we can obtain
\begin{widetext}
\begin{eqnarray}
\label{req}
&&[e^{-3\psi+\nu+\mu_{2}-\mu_{3}}(e^{3\psi+\nu-\mu_{2}-\mu_{3}}Q_{23}),_{3}],_{3}  + [e^{-3\psi+\nu+\mu_{2}+\mu_{3}}(e^{3\psi+\nu-\mu_{2}-\mu_{3}}Q_{23}),_{2}],_{2} = Q_{23,0,0}.
\end{eqnarray}
\end{widetext}
Eq. \eqref{req} can be separated by expanding the function in terms of Gegenbauer function $ \mathcal{C}_{n}^{\alpha}(\theta), $ that satisfies the following differential equation
\begin{equation}
\label{eq}
\left[\frac{d}{d\theta}\sin^{2\alpha}\theta\frac{d}{d\theta}+n(n+2\alpha)\sin^{2\alpha}\theta\right]\mathcal{C}_{n}^{\alpha}(\theta)=0.
\end{equation}
By defining the separable equation as
\begin{eqnarray}
e^{3\psi+\nu-\mu_{2}-\mu_{3}}Q_{23}= X(r) ~\mathcal{C}_{l+2}^{-3/2}(\theta),
\end{eqnarray}
we find 
\begin{eqnarray}
r^2 e^{\nu-\mu_{2}} \left( \frac{e^{\nu-\mu_{2}}}{r^2} X_{,r}\right)_{,r}-(l-1)(l+2) \frac{e^{2\nu}}{r^2}X+\sigma^2 X=0.\nonumber \\ 
\end{eqnarray}
Furthermore, by introducing the Tortoise coordinate
\begin{eqnarray}
\frac{dr_{*}}{dr} = e^{\mu_{2}-\nu},~~~\textrm{and}~~~ X(r)=r\Psi(r),
\end{eqnarray}
Eq. \eqref{req} can be reduced to the time-dependent wave equation in terms of $ x_{*}\equiv r_{*}/R $
\begin{equation}
\label{pdp}
\left[\frac{\partial^2}{\partial (t/R)^2}-\frac{\partial^2}{\partial x_{*}^2}+V(x)\right]\Psi(x_{*},t/R)=0,
\end{equation}
where the dimensionless effective radial potential, after adjusting the convention with our metric in \eqref{ansatz}, is 
\begin{widetext}
\begin{equation}
\label{veffint}
V(x)\equiv V(r) R^2 = \frac{e^{\nu}}{x^3} \bigg\lbrace l(l+1)x+ 4\pi x^3 \left[\rho\left(1+\frac{\chi}{4\pi} \right)-p \left(1+\frac{3\chi}{4\pi} \right) \right]   -6 m(x) \bigg\rbrace.
\end{equation}
\end{widetext}
The master equation above is the Regge-Wheeler equation. Inside the star, we use the effective interior potential expressed in \eqref{veffint}, whereas, outside the star, we use the Schwarzschild exterior equation.

\begin{figure*}
	\centering
	\includegraphics[width=0.47\textwidth]{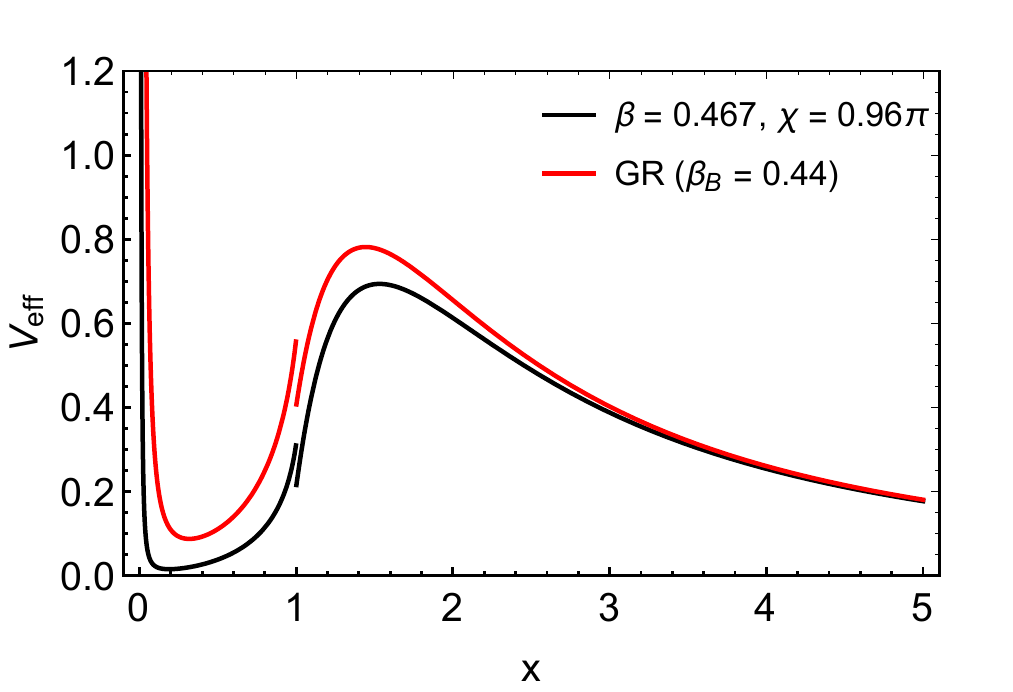}
	\includegraphics[width=0.47\textwidth]{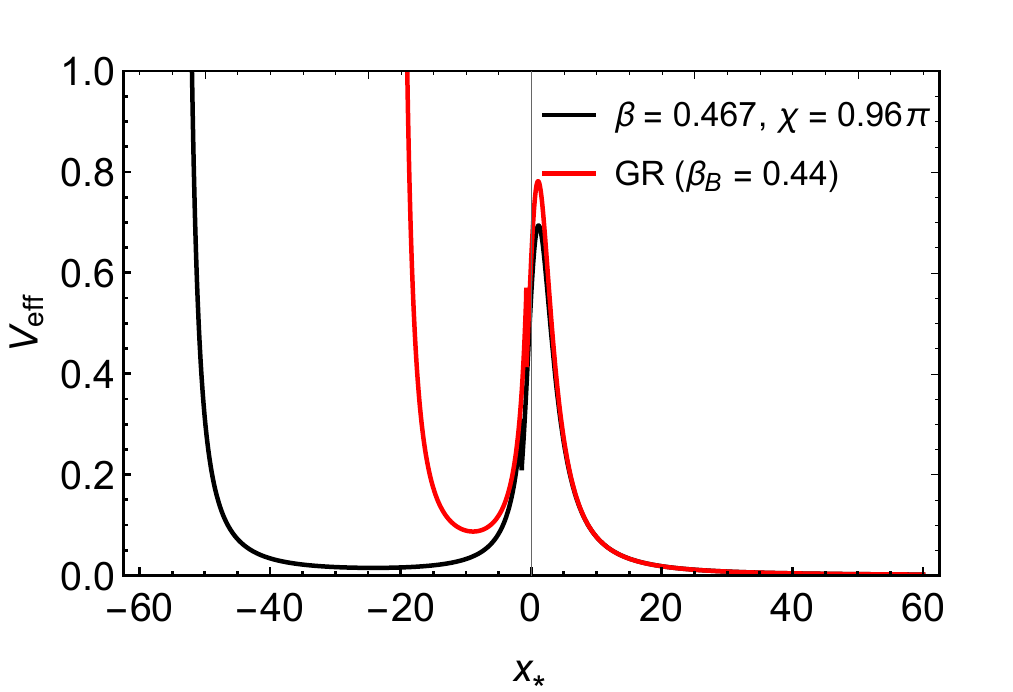}
	\caption{Dimensionless effective potential ($V_{\textrm{eff}}\equiv V(x)$) plots in terms of $x$ and tortoise coordinate $x_{*}$, respectively.}
	\label{fig:veff}
\end{figure*}

In Fig~\ref{fig:veff}, we show the dimensionless effective potential as a function of $x$ and tortoise coordinate through the equation $dx_{*}/dx=\sqrt{-g_{rr}/g_{tt}}$. Here, we compare the Buchdahl limit coming from GR and the Buchdahl limit from $f(R,T)$ gravity. Note that the coupling constant $\chi$ allows us to explore the compactness beyond standard GR, and the imprints of this aspect are the deeper potential well. At the surface ($x=1$), the results show a significant difference. The CDS EoS applied in both cases describe the constant density over the region inside the star, and abruptly goes to zero once they reach the surface. Therefore, the profile shows a discontinuity, and exhibits the potential well in the ultracompact object region. These signatures are important for the production of gravitational echoes  and QNMs. Moreover, the spectrum mode related to this kind of potential could contain the trapped modes that correspond to the oscillation generated by the ultracompact object star.

\begin{table*}[h!]
	\centering
	
	\begin{tabular}{|c|cc|cc||cc|cc|}
		\hline
		\multirow{3}{*}{$n$}
		& \multicolumn{4}{c||}{BSF}
		& \multicolumn{4}{c|}{BSL} \\ \cline{2-9}
		
		& \multicolumn{2}{c|}{$\Re(\omega_n M)$}
		& \multicolumn{2}{c||}{$|\Im(\omega_n M)|$}
		& \multicolumn{2}{c|}{$\Re(\omega_n M)$}
		& \multicolumn{2}{c|}{$|\Im(\omega_n M)|$} \\ \cline{2-9}
		
		& GR & FRTG & GR & FRTG
		& GR & FRTG & GR & FRTG \\
		\hline
		
		0 & 0.352 & 0.152 & $3.21\times10^{-7}$ & $1.13\times10^{-10}$
		& 0.241 & 0.151 & $2.29\times10^{-9}$ & $3.17\times10^{-11}$ \\ \hline
		
		1 & 0.494 & 0.223 & $1.09\times10^{-5}$ & $3.63\times10^{-9}$
		& 0.330 & 0.206 & $7.14\times10^{-8}$ & $8.76\times10^{-10}$ \\ \hline
		
		2 & 0.631 & 0.289 & $1.55\times10^{-4}$ & $3.84\times10^{-8}$
		& 0.420 & 0.262 & $9.33\times10^{-7}$ & $1.00\times10^{-8}$ \\ \hline
		
		3 & 0.758 & 0.349 & $1.32\times10^{-3}$ & $2.29\times10^{-7}$
		& 0.509 & 0.317 & $7.84\times10^{-6}$ & $7.28\times10^{-8}$ \\ \hline
		
		4 & 0.876 & 0.406 & $8.20\times10^{-3}$ & $9.81\times10^{-7}$
		& 0.598 & 0.373 & $5.14\times10^{-5}$ & $3.96\times10^{-7}$ \\ \hline
		
		5 & 0.989 & 0.459 & $4.12\times10^{-2}$ & $3.37\times10^{-6}$
		& 0.686 & 0.429 & $3.20\times10^{-4}$ & $1.78\times10^{-6}$ \\ \hline
		
		6 & 1.109 & 0.509 & $1.75\times10^{-1}$ & $9.95\times10^{-6}$
		& 0.771 & 0.484 & $1.24\times10^{-3}$ & $8.88\times10^{-6}$ \\ \hline
		
		7 & & 0.557 & & $2.62\times10^{-5}$
		& 0.855 & 0.539 & $4.52\times10^{-3}$ & $2.59\times10^{-5}$ \\ \hline
		
		8 & & 0.603 & & $6.33\times10^{-5}$
		& & 0.593 & & $7.11\times10^{-5}$ \\ \hline
		
		9 & & 0.648 & & $1.43\times10^{-4}$
		& & 0.648 & & $1.92\times10^{-4}$ \\ \hline
		
		10 & & 0.691 & & $3.04\times10^{-4}$
		& & 0.703 & & $4.99\times10^{-4}$ \\ \hline
		
		11 & & 0.733 & & $6.17\times10^{-4}$
		& & 0.757 & & $1.25\times10^{-3}$ \\ \hline
		
		12 & & 0.774 & & $1.21\times10^{-3}$
		& & 0.811 & & $2.95\times10^{-3}$ \\ \hline
		
	\end{tabular}
	
	\vspace{0.4cm}
	
	\begin{tabular}{|c|cc|cc|}
		\hline
		\multirow{3}{*}{$n$}
		& \multicolumn{4}{c|}{BSH} \\ \cline{2-5}
		
		& \multicolumn{2}{c|}{$\Re(\omega_n M)$}
		& \multicolumn{2}{c|}{$|\Im(\omega_n M)|$} \\ \cline{2-5}
		
		& GR & FRTG & GR & FRTG \\
		\hline
		
		0 & 0.247 & 0.154 & $2.91\times10^{-9}$ & $4.01\times10^{-11}$ \\ \hline
		1 & 0.336 & 0.209 & $8.41\times10^{-8}$ & $1.03\times10^{-9}$ \\ \hline
		2 & 0.425 & 0.258 & $1.07\times10^{-6}$ & $8.57\times10^{-9}$ \\ \hline
		3 & 0.514 & 0.301 & $8.77\times10^{-6}$ & $4.21\times10^{-8}$ \\ \hline
		4 & 0.626 & 0.375 & $1.16\times10^{-5}$ & $4.14\times10^{-7}$ \\ \hline
		5 & 0.691 & 0.459 & $3.48\times10^{-4}$ & $3.90\times10^{-6}$ \\ \hline
		6 & 0.780 & 0.486 & $1.45\times10^{-3}$ & $9.12\times10^{-6}$ \\ \hline
		7 & 0.846 & 0.524 & $3.98\times10^{-3}$ & $1.97\times10^{-5}$ \\ \hline
		8 & & 0.604 & & $8.67\times10^{-5}$ \\ \hline
		9 & & 0.649 & & $1.94\times10^{-4}$ \\ \hline
		10 & & 0.703 & & $4.95\times10^{-4}$ \\ \hline
		11 & & 0.756 & & $1.21\times10^{-3}$ \\ \hline
		12 & & 0.807 & & $2.77\times10^{-3}$ \\ \hline
		
	\end{tabular}
	
	\caption{Numerical values of the quasinormal modes for the Bohr--Sommerfeld fitting (BSF), Bohr--Sommerfeld low-order (BSL), and Bohr--Sommerfeld high-order (BSH) approximations with $l=2$.}
	\label{table:QNM_values}
\end{table*}

QNM, in general, is a dissipative property of spacetime and plays a crucial role in the form of gravitational wave echoes when the stars reach the final stage during a ringdown phase. In this work, we follow the procedure in Ref.~\cite{Volkel:2017ofl} to obtain the QNM mode using the Wentzel-Kramers-Brillouin (WKB) approximation. In quantum mechanics, the Bohr-Sommerfeld (BS) rule is a well-known method to obtain the approximate energy spectrum $ E_{n}, $ of bound states in a potential. It is possible to include higher-order corrections of the BS rule by utilizing WKB theory~\cite{popov}. It means that WKB approximation is the generalized version of the BS rule \cite{atomicwkb},
\begin{equation}
\label{BS}
\int_{x_{0}}^{x_{1}} \sqrt{E_{n}-V(x)} dx = \pi \left(n + \frac{1}{2}\right) - \frac{i}{4} e^{2i\int_{x_{1}}^{x_{2}} \sqrt{E_{n}-V(x)} ~dx},
\end{equation}
where $ x_{0} $ and $ x_{1} $ are the classical turning points determined by the root of the integrand and also depend on the energy spectrum. The second term in Eq.~\eqref{BS} is the additional term described as the general one for the BS rule, where $ x_{2} $ is the third classical turning point on the right side of the potential barrier. For a detailed discussion of this procedure, please see Ref.~\cite{Volkel:2017ofl}. This additional term denotes an exponentially small imaginary part of the energy spectrum, which measures the barrier penetrability. We can simplify further by writing the energy spectrum as $ E_{n}=E_{0n}+iE_{1n} $, where $ E_{0n} $ is the real part of the energy and $ E_{1n} $ is the small imaginary energy. Substituting these energy spectra into the left-hand side of Eq.~\eqref{BS} and matching with the real and imaginary part on the right-hand side of Eq.~\eqref{BS} yields
\begin{equation}
\label{e0n}
\int_{x_{0}(E_{0n})}^{x_{1}(E_{0n})} \sqrt{E_{0n}-V(x)} dx = \pi \left(n + \frac{1}{2}\right),
\end{equation}
and
\begin{widetext}
\begin{eqnarray}
\label{e1n}
E_{1n} &=& -\frac{1}{2}\left( \int_{x_{0}(E_{0n})}^{x_{1}(E_{0n})} \frac{1}{\sqrt{E_{0n}-V(x)}} dx \right)^{-1}  e^{2i \int_{x_{1}(E_{0n})}^{x_{2}(E_{0n})} \sqrt{E_{0n}-V(x)} ~ dx}.
\end{eqnarray}
\end{widetext}
Through root-finding and integration, we can numerically solve Eq.~\eqref{e0n} and have the resulting $E_{0n}$ be substituted in Eq.~\eqref{e1n} to obtain $E_{1n}$. This method is referred to as the low order Bohr-Sommerfeld rule \cite{Volkel:2017ofl}.

It is possible to improve the accuracy in calculating the real component of the energy $E_{0n}$ by incorporating higher order corrections to Eq.~\eqref{e1n}. This results in the high order Bohr-Sommerfeld rule, with the form \cite{atomicwkb, Volkel:2017ofl}
\begin{widetext}
\begin{equation}
\label{eqgav:BSH}
\int_{x_{0}(E_{0n})}^{x_{1}(E_{0n})} \sqrt{E_{0n}-V(x)} dx = \pi \left(n + \frac{1}{2}\right) +\frac{1}{24}\frac{\partial^2}{\partial E^2_{n}}\int_{x_{0}(E_{0n})}^{x_{1}(E_{0n})} \left( \frac{dV(x)}{dx}\right)^2 \frac{dx}{\sqrt{E_{0n}-V(x)}}.
\end{equation}
\end{widetext}
Alternatively, we can take a different approach on Eqs.~\eqref{e0n}~and~\eqref{e1n} by fitting the potentials with simpler functions that can be solved analytically. We use the well-known analytic fitting potential described below \cite{Volkel:2017ofl}
\begin{eqnarray}
U_{Q} = U_{0}+\lambda_{0}^2 (x-x_{min})^{4},
\end{eqnarray}
and
\begin{eqnarray}
 U_{BW} = \frac{U_{1}}{1+\lambda_{1}(x-x_{max})^2}.
\end{eqnarray}   
The functions above are called the quartic oscillator and the Breit-Wigner potentials, respectively. These functions depend on the $ U_{0}, $ $ U_{1}, $ $ \lambda_{0}, $ and $ \lambda_{1}$ parameters. Then, the value of these parameters is matched with the true effective potential (in tortoise coordinate). The variable $ U_{0} $ will be the interior of the true potential, whereas the variable $ U_{1} $ is the one for the maximum of the true potential. Next, $ \lambda_{1} $ can be obtained by identifying $ V_{max}'' $ on the $ U_{BW} $ function, which can be written as $ \lambda_{1}=-V_{max}''/2V_{max}. $ The last parameter, $ \lambda_{0}$, demands that the quartic oscillator must be equal to the Regge-Wheeler equation at the surface. However, we can start to evaluate the energy spectrum by inserting both fitting functions into Eq.\eqref{e0n} and Eq.\eqref{e1n}. The solution reads
\begin{widetext}
\begin{eqnarray}
\label{E0n}
E_{0n} &=& U_{0} + \lambda_{0}^{2/3} \left[\frac{3\pi}{4~ \mathcal{K}(-1)} \left(n+\frac{1}{2}\right)\right]^{4/3}, \\ E_{1n} &=& -\frac{\sqrt{\lambda_{0}}(E_{0n}-U_{0})^{1/4}}{4~ \mathcal{K}(-1)} ~\textrm{Exp}\bigg[4i  \sqrt{\frac{U_{1}-E_{0n}}{\lambda_{1}}}~  \mathcal{E}\left( i \sinh^{-1}\left(\sqrt{\frac{U_{1}}{E_{0n}}-1}\right),\frac{E_{0n}}{E_{0n}-U_{1}} \right)   \bigg],\nonumber \\
\label{tuti}
\end{eqnarray}
\end{widetext}
where $ \mathcal{K}(a) $ is the complete elliptic integral of the first kind and $ \mathcal{E}(a,b) $ denotes the elliptic integral of the second kind \cite{Abraham}. The variable $ n $ shown in Eq.~\eqref{E0n} is the overtone number. We can match the energy with the complex eigenvalue which can be written as $ \omega_{n}=\sqrt{E_{n}}=\sqrt{E_{0n}+iE_{1n}}=\omega_{r}+i\omega_{i}$, where the real part describes the frequency of the oscillation and the imaginary part describes the inverse of the damping time, $ \tau_{d} $. This brings us to the approximate analytic results for all trapped modes.

The exact values of the QNMs are presented in TABLE~\ref{table:QNM_values}. The frequencies in TABLE~\ref{table:QNM_values} is calculated using the various Bohr-Sommerfeld methods as discussed before, where the "BSF" column uses the Bohr-Sommerfeld Fitting approach (Eqs.~\eqref{E0n}~and~\eqref{tuti}), the "BSL" column uses the low order Bohr-Sommerfeld rules (Eqs.~\eqref{e0n}~and~\eqref{e1n}), and the "BSH" column uses the high order Bohr-Sommerfeld rules (Eqs.~\eqref{e0n}~and~\eqref{eqgav:BSH}). This discussion focuses on the QNMs frequencies of both the FRTG and GR models with the compactness of each one's Buchdahl limit (Eq.~\eqref{betabuchdal}). It can be seen that, compared to the GR model, the QNMs of the FRTG model have lower frequencies and lower damping rates. Furthermore, the number of available overtone numbers is increased for the FRTG model compared to the GR model. Both of these phenomena are caused by the widening of the Regge-Wheeler potential well where the well floor is lowered and its width widened.

\begin{figure}[h!]
	\centering
	\includegraphics[width=0.47\textwidth]{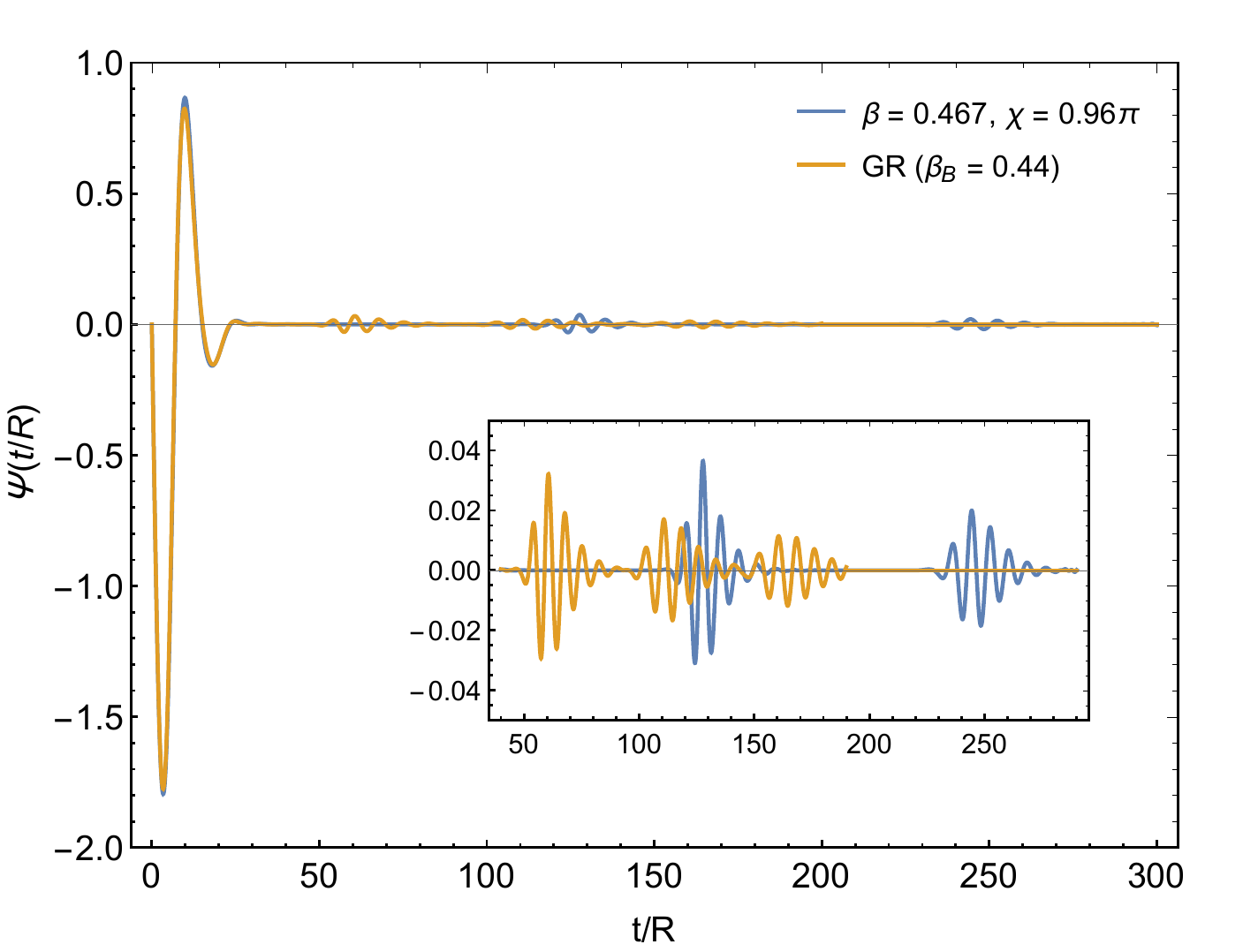}
	\includegraphics[width=0.45\textwidth]{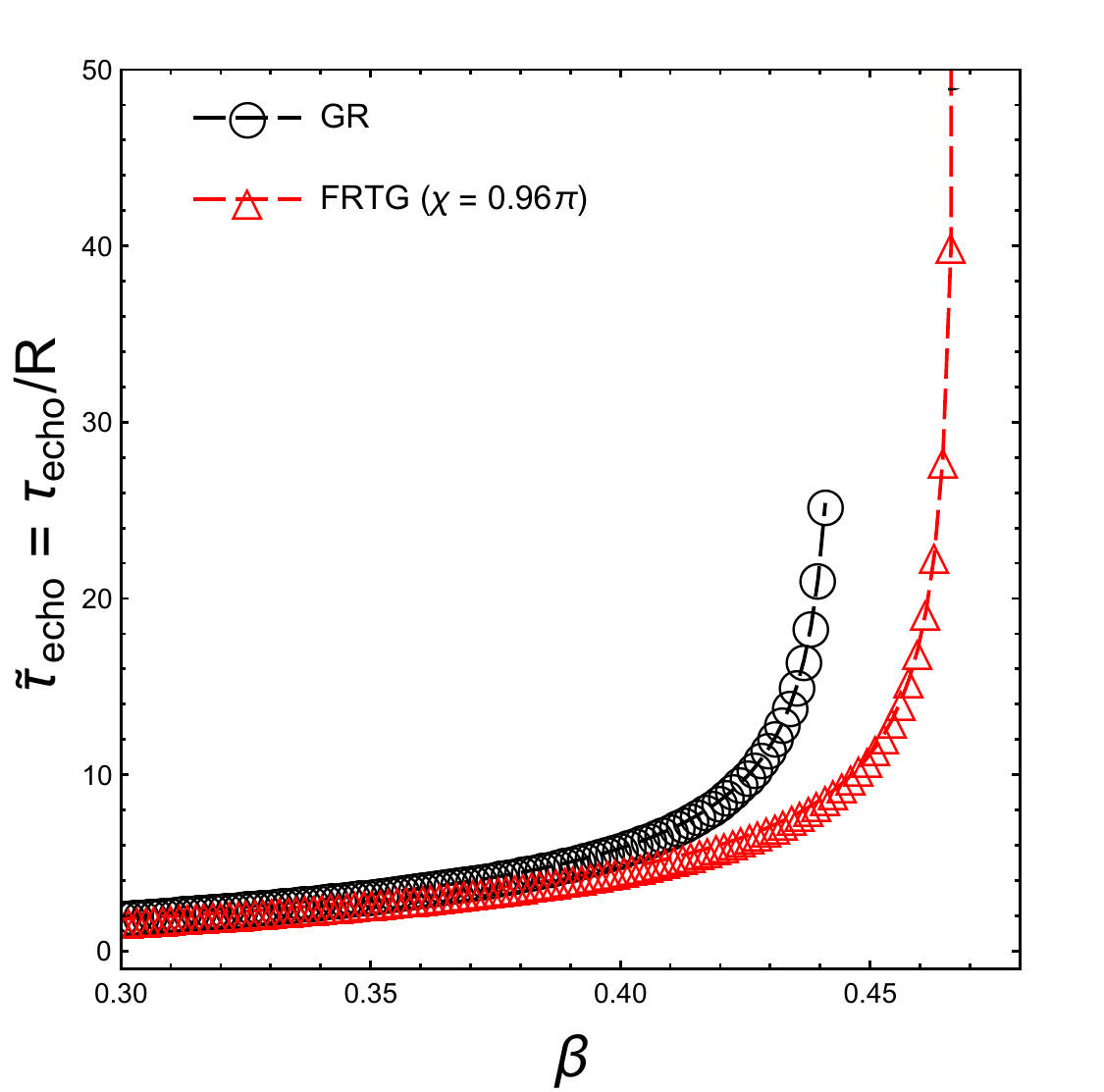}
	\caption{[Left] The gravitational wave echoes vs $t/R$. [Right] The echo time ($\tilde{\tau}$) vs compactness ($\beta$). In the gravitational-wave echo plot, observe that the train echoes increase when the stellar object is less compact. In the echo time plot, we set the lower bound to be the light ring, where the potential well starts to appear.}
	\label{fig:gravecho}
\end{figure}

Next, we numerically analyze the gravitational echoes from the perturbed ultracompact star by solving the time-dependent partial differential equation in Eq.~\eqref{pdp}. The solution requires four conditions, 
\begin{eqnarray}
\textrm{I.}&&~\Psi(r_{*},0)~~~~~~~~~~=~0,\nonumber\\
\textrm{II.}&&~\Psi(r_{*}^{c},t)~~~~~~~~~~=~0,\nonumber\\ 
\textrm{III.}&&~\frac{\partial \Psi(t,r_{*})}{\partial t}\bigg|_{t=0}~~~=~f(r_{*}),\nonumber\\
\textrm{IV.}&&~\frac{\partial \Psi(r_{*},t)}{\partial r}\bigg|_{r_{*}\rightarrow \infty} = -\frac{\partial \Psi(r_{*},t)}{\partial t}\bigg|_{r_{*}\rightarrow \infty}.
\end{eqnarray}
Conditions I and III are the initial data, and II and IV are the boundary conditions. Both initial data denotes the post-merger phase with an initial Gaussian pulse centered at $ r_{*}=r_{g} $ and with spread $ \sigma $; $ f(r_{*})=\textrm{Exp}\big[-(r_{*}-r_{g})^2/\sigma^2\big] $ \cite{Urbano:2018nrs,Cardoso:2016oxy}. III. is the condition for the star to be regular at the center, where $ r_{*}^{c}(r)=r_{*}(0)$. Lastly, condition IV relates outgoing waves at spatial infinity \cite{Urbano:2018nrs}. 

After evaluating numerically, the solution of the differential equation depends on the time and tortoise radius, $ \Psi(x_{*},t/R). $ In this discussion, we focus on the time evolution of the signal and present only the lowest $(l=2)$ mode, $\Psi(x_{*}=5,t/R)$. The signal can be obtained by inserting the effective potential (as a function of tortoise coordinate $ x_{*} $) into Eq.~\eqref{pdp}, and the results are presented in Fig.~\ref{fig:gravecho}. In the initial condition, we impose the same Gaussian injection value, and the discrepancy occurs in the late-time evolution. The train wave after the merger state represents the signal entering the interior and then getting reflected by the center of the star. Therefore, we can infer that the FRTG model enables the existence of a less-train wave than the GR counterpart. It can be physically expected due to the wider potential well, and therefore the wave travels longer inside the star. The round-trip wave can also be obtained by studying the echo time. The expression can be written in the dimensionless form as follows
\begin{equation}
\tau_{\textrm{echo}} = \int_{0}^{3M} \sqrt{-\frac{g_{rr}}{g_{tt}}} dr 
\end{equation}
where the metric in the integral is the complete set of the solution from the interior to the exterior region. The metric components $tt$ and $rr$ in the right side are expressed in the dimensionless form $x$. We can see the numerical results in the right panel of Fig.~\ref{fig:gravecho}, where the echo time is plotted against the compactness. The generic profile is that the higher the compactness, the longer the echo time because the potential well is wider. In our particular case, the dimensionless echo time ($\tilde{\tau}$) of the maximum compactness of each model (with $\chi=0.96\pi$ for the FRTG model) are $\tilde{\tau}_{\textrm{echo}} =6.29\times 10^{3} $ (FRTG), and $\tilde{\tau}_{\textrm{echo}} =2.53\times 10^{1} $ (GR).


However, under the proposed gravitational-wave scenario, the existence of a photon sphere is required for the star to support perturbations propagating with the characteristic quasinormal-mode frequency. In the $f(R,T)$ model, we consider $\chi=0.96\pi$, for which the allowed range of compactness is $\beta\in[10^{-3},0.4679]$. Moreover, part of this range extends beyond the photon-sphere threshold, $\beta=1/3$.

For GW170817, the estimated echo frequency \cite{Mannarelli:2018pjb} is defined as
\begin{equation}
f_{\textrm{echo}}=\frac{1}{2\tau_{\textrm{echo}}},
\end{equation}
where $\tau_{\textrm{echo}}$ denotes the echo time delay. The analysis of GW170817 reported a characteristic frequency of approximately $72$ Hz with a significance of $4.2\sigma$ \cite{Mannarelli:2018pjb}. We therefore compare this observationally inferred frequency with the prediction obtained from our $f(R,T)$ model and the GR case, assuming the object is an NS.

We first adopt the stellar radius constrained by the Neutron Star Interior Composition Explorer (NICER) \cite{Dittmann:2024mbo}, $R=11.4$ km. For $\beta=0.4662702$ and $\chi=0.96\pi$, we obtain an echo time of
\begin{equation}
\tau_{\textrm{echo}}=7.18\times10^{4}\ {\rm km}.
\end{equation}
Converting to physical units using geometrized units, this corresponds to
\begin{equation}
\tau_{\textrm{echo}}=2.39\times10^{2}\ {\rm ms},
\end{equation}
and consequently gives
\begin{equation}
f_{\textrm{echo}}=72.11\ {\rm Hz}.
\end{equation}
This value is remarkably close to the observationally inferred frequency of $72$ Hz. In contrast, for the GR case, we obtain $f_{\textrm{echo}}=5.19\times10^{2}$ Hz, which is substantially higher than the reported value. Therefore, our results indicate that the $f(R,T)$ model considered here can produce an echo frequency that is consistent with the characteristic frequency inferred from GW170817, whereas the GR case does not exhibit the same level of agreement.

\section{Conclusion}
\label{section5}

We constructed the CDS in $f(R,T)=R+2\chi T$ gravity. The field equation in this model exhibits the term $\delta^2\mathcal{L}_{m}/\delta g^{\mu\nu}\delta g^{\alpha\beta}$, which can be chosen as either $\mathcal{L}_{m}=p$ or $\mathcal{L}_{m}=-\rho$. In \cite{Pappas:2022gtt}, the author used the same exact model with $\mathcal{L}_{m}=p$, while our work has a different scenario in choosing the $\mathcal{L}_{m}$ (see the details in \cite{Jayawiguna:2026das}). Therefore, we have different metric tensor solutions from \cite{Pappas:2022gtt}. In the parametric $\chi-\beta$ space, where $\beta$ is the compactness of the star, we found that the FRTG model can have bounded values in certain ranges of $\chi$; $\chi\in[-8\pi,16\pi]$ for Buchdal limit, $\chi \in[-8\pi,4\pi]$ for the subluminal ($0<c_{s}^2<1$) condition, and $\chi \in[-4\pi,12\pi]$ for dominant energy condition (DEC) value. The special properties happen in $\chi=4\pi,$ where the Buchdahl profile coincides with the DEC limit at the BH compactness limit.

We extend our study to consider the observational imprint through the axial perturbation  of the CDS model. In $\chi>0,$ the Buchdahl limit in $f(R,T)$ gravity is higher than GR, which allows us to set $\beta >0.44$. We found that the {effective potential} with maximum compactness has a smaller well in GR that in $f(R,T)$ gravity. It indicates a shorter echo time since the light travels faster in the smaller potential. The characteristic frequency, or QNM, in both cases (GR and FRTG) lies in the ultracompact object regime and hence the trapped mode will propagate through the barrier. In obtaining the QNM, we have used three different Bohr-Sommerfeld methods based on WKB approximations, that is BSF, BSL, and BSH \cite{Volkel:2017ofl}. We have found that for each overtone number, the GR QNM frequency is consistently higher than their FRTG counterpart, indicating that it has higher frequencies while also having a shorter damping time. We have also found that the FRTG model allows for higher available overtone numbers compared to the GR model as a result of the widened potential well. Finally, for $f(R,T)$ gravity with $\chi=0.96\pi$ and $\beta=0.4662702$, the characteristic frequency is consistent with GW170817 ($f\approx 72 $ Hz).

All in all, we have investigated the extended Tolman III within the $f(R,T)$ gravity theory, including the axial perturbation sector. In \cite{Pappas:2022gtt}, the authors include the extended version of the Tolman VII model, which is interesting to see the impact of the new formalism to the parametric space and the perturbation properties. We leave this for further investigations. On the other hand, the topic we discussed is a spherically symmetric object, which is an idealized object. In reality, the unavoidable rotation breaks the symmetry and represents a real case. As a starting point, we can assume that the star rotates slowly, so we can expand the spherical harmonics up to second-order and make it easier to solve. The corresponding exterior properties under this rotational perturbation such as moment of inertia, Love number, and the quadrupole deformation, are a focal point to investigate the multipole moments properties of the object. These promising ideas could be explored in future work.


\acknowledgments

We thank Avrill Aiko Lyxie, Fahmi Fauzi and Muhammad Hafizt Al Ghifari for the useful discussions. BNJ and AS are supported by Hibah Fundamental DIKTI, PKS-178/UN2.RST/HKP.05.00/2026. The authors acknowledge the use of ChatGPT (OpenAI) for assistance with language editing and grammatical refinement of the manuscript. The authors take full responsibility for the content of the manuscript. \\

\section*{Statement of the Conflict of Interest}

All authors hereby declare that we do not have any conflict of interest.

\section*{Data Availability Statement}

Data sharing is not applicable to this article as no data sets were generated or analyzed during the current study.





\nocite{*}


\end{document}